\documentclass[12pt]{cernrep}
\usepackage{graphicx}

\newcommand{\Pt}{{P_t}}
\newcommand{\dphi}{\Delta\phi}
\newcommand{\phigj}{\phi_{(\gamma,jet)}}

\newcommand{\gpj}{~"$\gamma+Jet$"~}
\newcommand{\rrr}{\to} 

\newcommand{\pth}{\hat{p}_{\perp}^{\;min}}

\newcommand{\ptgj}{$~\Pt^{\gamma}$ and $\Pt^{Jet}~$}
\newcommand{\ptg}{$\Pt^{\gamma}$}
\newcommand{\Ptg}{\Pt^{\gamma}}

\newcommand{\Gvc}{\footnotesize{$(GeV/c)$} }

\newcommand{\hmm}{\hspace*{-1.3mm}}

\begin{document}
\thispagestyle{empty}
\thispagestyle{empty}

\vskip-5mm

\begin{center}
{\Large JOINT INSTITUTE FOR NUCLEAR RECEARCH}
\end{center}

\vskip10mm

\begin{flushright}
Subm. to ``Physics of \\
Particles and Nuclei Letters''\\
hep-ex/0011015
\end{flushright}

\vspace*{3cm}

\begin{center}
\noindent
{\Large{\bfseries \gpj events rate estimation for gluon distribution determination at LHC.}}
\\[5mm]
{\large D.V.~Bandourin$^{1\,\dag}$, V.F.~Konoplyanikov$^{2\,\ast}$, N.B.~Skachkov$^{3\,\dag}$}

\vskip 0mm

{\small
{\it
E-mail: (1) dmv@cv.jinr.ru, (2) kon@cv.jinr.ru, (3) skachkov@cv.jinr.ru}}\\[3mm]
$\dag$ \large \it Laboratory of Nuclear Problems \\
\hspace*{-4mm} $\ast$ \large \it Laboratory of Particle Physics
\end{center}

\vskip 9mm
\begin{center}
\begin{minipage}{150mm}
\centerline{\bf Abstract}
\noindent
It is shown that \gpj events, being collected at LHC,
would provide us with the data sufficient
for an extraction of gluon distribution function in a proton using valence and sea
quark distributions measured in the same experiment with another physical processes.
A new region of $10^{-4}\leq x \leq 10^{-1}$ with
$1.6\cdot10^3\leq Q^2\leq10^5 ~(GeV/c)^2$ can be covered.
The rates of $g\,c\to \gamma^{dir} \,+\,Jet$ events are also given. 
\end{minipage}
\end{center}

\newpage

\section{INTRODUCTION}
As many of theoretical predictions  for new particles
(Higgs, SUSY) production at LHC are based
on model estimations of gluon density behavior at
low $x$ and high $Q^2$, the measurement of proton gluon density
for this kinematical region directly in LHC experiments
would be obviously quite useful. One of the promising
channels for this measurement, as it was shown in ~\cite{Au1},
is a high $\Pt$ direct photon production $pp\rightarrow \gamma^{dir} + X$.
The region of high $\Pt$,
reached up to now by UA1 \cite{UA1}, UA2 \cite{UA2}, CDF \cite{CDF1} and 
D0 \cite{D0} extends up to
$\Pt \approx 60~ GeV/c$. These data together with the latter ones
(see references in \cite{Fer}--\cite{Fr1}) and recent 
E706 \cite{E706} and UA6 \cite{UA6} results give
an opportunity for tuning the form of gluon distribution (see \cite{Au2},
\cite{Vo1}, \cite{Mar}).
The rates and estimation for cross sections of inclusive
direct photon production at LHC are given in [1]
(see also \cite{AFF}). 

Here we shall consider  another process (see also \cite{MD1})\\[-19pt]
\begin{eqnarray}
pp\rightarrow \gamma^{dir}\, +\, 1\,Jet\, + \,X
\label{1}
\end{eqnarray}

\noindent
(for experimental results see \cite{ISR}, \cite{CDF2})
that at the leading order is defined by two QCD subprocesses:
``Compton-like" process (which gives the main contribution)~ $q g\rrr \gamma +q$ and
``annihilation'' process~ $q \bar{q}\rrr \gamma + g$~.

The study of $\gamma^{dir} + 1 \,Jet$ final state is a more preferable one
from the viewpoint of extraction of information on gluon distribution.
In the case of inclusive direct photons production
the cross section is given as an intergral over partons (a,b = quarks and gluon)
distribution functions $f_a(x_a,Q^2)$, while for (1) at $\Pt\, \geq \,30\, GeV/c$
(i.e. the region where $k_T$ smearing effects should not be important, see \cite{Hu2})
it is expressed directly
through these distributions (see, for example, \cite{Owe}; $\eta_1=\eta^\gamma$,
$\eta_2=\eta^{jet}$; ~$\Pt=\Pt^\gamma$;~ a,b = $q, \bar{q},g$; 3,4 = $q,\bar{q},g,
\gamma$)\\[-15pt]
\begin{eqnarray}
\frac{d\sigma}{d\eta_1d\eta_2d\Pt^2} = \sum\limits_{a,b}\,x_a\,f_a(x_a,Q^2)\,
x_b\,f_b(x_b,Q^2)\frac{d\sigma}{d\hat{t}}(a\,b\rightarrow 3\,4)
\label{4}
\end{eqnarray}
\noindent
where $x_{a,b} \,=\,\Pt/\sqrt{s}\cdot \,(exp(\pm \eta_{1})\,+\,exp(\pm \eta_{2}))$.
Formula (4) with the knowledge of the results of independent measurements of
$q, \,\bar{q}$ distributions \cite{MD1} allows to determine gluon $f_g(x,Q^2)$ 
distribution.

Our work is based on the results of ~\cite{BKS}, where the selection criteria of
\gpj events
with a clean topology and most suitable for jet energy absolute scale setting at LHC
energy were developed. In \cite{BKS} mainly PYTHIA was used comlemented by GEANT simulation
to study a possibility of the background events rejection.
Below the CMS detector geometry will be used as an example.
\section{DEFINITION OF SELECTION RULES.}
Our selection conditions for \gpj events were
chosen as in \cite{BKS}. We suppose the ECAL size to be limited by
$|\eta| \leq 2.61$ and HCAL is limited by $|\eta| \leq 5.0$ (CMS geometry),
where $\eta = -ln (tan (\theta/2))$ is a pseudorapidity defined through
a polar angle $\theta$ counted from the beam line. In the plane
transverse to the beam line the azimuthal angle $\phi$ defines the
directions of $\vec{\Pt}^{Jet}$ and $\vec{\Pt}^{\gamma}$.

\noindent
1. We select the events with one jet and one photon candidate with\\[-5pt]
\begin{equation}
\Pt^{\gamma} \geq 40 \; GeV/c~; \qquad  \Pt^{jet} \geq 30 \;GeV/c.
\label{eq:sc1}
\end{equation}
The jet is defined here according to PYTHIA jetfinding algorithm LUCELL.
The jet cone radius R in $\eta-\phi$ space is
taken as $R=((\Delta\eta)^2 + (\Delta\phi)^2)^{1/2}=0.7$.

\noindent
2. To suppress the background processes, only the events with "isolated"
photons are taken. To do this, we restrict:

a) the value of the scalar sum of $\Pt$ of hadrons and other particles surrounding
a photon within a cone of $R^{\gamma}_{isol}=( (\Delta\eta)^2 + (\Delta\phi)^2)^{1/2}=0.7$
(``absolute isolation cut")\\[-7pt]
\begin{equation}
\sum\limits_{i \in R} \Pt^i \equiv \Pt^{isol} \leq \Pt_{CUT}^{isol};
\label{eq:sc2}
\end{equation}
\vspace{-2.6mm}

b) the value of a fraction (``relative isolation cut'')\\[-7pt]
\begin{equation}
\sum\limits_{i \in R} \Pt^i/\Pt^{\gamma} \equiv \epsilon^{\gamma} \leq 
\epsilon^{\gamma}_{CUT};
\label{eq:sc3}
\end{equation}

c) we accept only the events having no charged tracks (particles) 
with $\Pt>1~GeV/c$ within the $R^{\gamma}_{isol}$ cone around the photon candidate.

\noindent
3. To be consistent with the application condition of the NLO
formulae, one should avoid an infrared dangerous region and take care of
$\Pt$ population in the region close to a photon (see  \cite{Ber}--\cite{FFT}).
In accordance with  \cite{Fr2} we also restrict the scalar sum of $\Pt$ of particles
 around a photon within a cone of a smaller radius $R_{singl} = 0.175 = 1/4
\,R_{isol}^{\gamma}$.

Due to this cut,
\begin{equation}
\sum\limits_{i \in R_{singl}} \Pt^i \equiv \Pt^{singl} \leq 2~ GeV/c,
~~~~~(i\neq ~\gamma-dir).
\label{eq:sc4}
\end{equation}
an ``isolated" photon with high $\Pt$ also becomes ``single'' within
an area of 8 towers (of 0.087x0.087 size according to CMS geometry)
 which surround the tower hitted by
it (analog of 3$\times$3 tower window algorithm).

\noindent
4. We also consider the structure of every event with the photon
candidate at a more precise level of 5x5 crystal cells window (size of one CMS
HCAL tower) with a cell size of 0.0175x0.0175. To suppress the background
events with photons resulting from high energetic $\pi^0-$, $\eta-$, $\omega-$
and $K_S^0-$ mesons,we apply in addition the following cuts:

a) the ECAL signal can be considered as a candidate to be a direct photon if it fits
inside the 3x3 ECAL crystal cell window (typical size of photon shower
in ECAL found from GEANT simulation with CMSIM package) 
with the highest $\Pt$ of $\gamma/e$ in the center;

b) the value of a scalar sum of $\Pt$ ($\Pt^{sum}$) of
stable particles in the 5x5 crystal cell window in the region
out of a smaller 3x3 crystal cell window having the cell
with the direct photon candidate (i.e. with the largest $\Pt$ of $\gamma/e$)
as the central one,
should be restricted by $1 \, GeV/c$, i.e.
\begin{equation}
\Pt^{sum} \leq 1~ GeV/c;
\label{eq:sc6}
\end{equation}

c) we require the absence of a high $\Pt$ hadron
in this 5x5 crystal cell window (that means an imposing of an upper cut
on the HCAL signal at least in the one-tower area) around the direct photon:
\begin{equation} 
\Pt^{hadr} \leq 5~ GeV/c. 
\label{eq:sc5} 
\end{equation}

We can not reduce this value to, for example, 2-3 $GeV/c$, because
a hadron with $\Pt$ below 2-3 $GeV/c$ deposits most of its energy in ECAL and
may not reveal itself in HCAL.

\noindent   
5. The events with the vector $\vec{\Pt}^{Jet}$ being ``back-to-back" to
the vector $\vec{\Pt}^{\gamma}$ in the transverse to a beam line plane
within $\dphi$ which is defined by equation:\\[-5pt]
\begin{equation}
\phigj=180^\circ \pm \Delta\phi \quad (\Delta\phi =15^\circ, 10^\circ, 5^\circ)
\label{eq:sc7}
\end{equation}
($5^\circ$ is a size of one CMS HCAL tower in $\phi$)
for the following definition of the angle $\phigj$: \\
\hspace*{2cm} $\vec{\Pt}^{\gamma}\vec{\Pt}^{Jet}=\Pt^{\gamma}\Pt^{Jet}\cdot cos(\phigj)$, ~~~
with ~$\Pt^{\gamma}=|\vec{\Pt}^{\gamma}|,~~\Pt^{Jet}=|\vec{\Pt}^{Jet}|$.
\noindent   

\noindent
6. To discard the background events,
we choose the events that do not have any other (exept one jet)
minijet-like or cluster high $\Pt$ activity (taking
the cluster cone $R_{clust}(\eta,\phi)=0.7$) with the $\Pt^{clust}$
higher than some threshold $\Pt^{clust}_{CUT}$ value. Thus 
we select events with\\[-5pt]
\begin{equation}
\Pt^{clust} \leq \Pt^{clust}_{CUT},
\label{eq:sc8}
\end{equation}
where clusters are found by the same jetfinder (LUCELL) used to find the jet
in the event.

\noindent
7. We limit the value of modulus of the vector sum of $\vec{\Pt}$ of all
particles that fit into the region covered by
ECAL and HCAL except the  \gpj system
(i.e. the cells ``out of the jet and photon'' regions):
\begin{equation}
\left|\sum_{i\not\in jet,\gamma-dir}\vec{\Pt}^i\right| \equiv \Pt^{out} \leq \Pt^{out}_{CUT}, ~~|\eta|<5
\label{eq:sc9}
\end{equation}

\noindent
8. To reduce the value of $\Pt^{Jet}$ uncertainty due to possible presence of
the neutrino
contribution to a jet and to diminish background events with 
high energetic electrons  \cite{BKS}, we select only events with a small 
$\Pt^{miss}$ value:\\[-21pt]
\begin{eqnarray}
\Pt^{miss}~\leq \Pt^{miss}_{CUT}.
\label{eq:sc11}
\end{eqnarray}

\noindent
9. In addition to selection cuts 1 -- 8 one more
new object, named an ''isolated jet'', will be introduced.
To do this, we also involve a new requirement of ``jet isolation'',
i.e. the presence of a ``clean enough'' (in the sense of small $\Pt$
activity) region inside the ring (of $\Delta R=0.3$ size) around the
jet.  Following this picture we restrict the value of the ratio of scalar sum
of particles transverse momenta belonging to this ring, i.e.\\[-5pt]
\begin{equation}
\Pt^{ring}/\Pt^{\gamma} \equiv \epsilon^{jet} \leq 2\%
, \quad {\rm where ~~~~ }
\Pt^{ring}=\sum\limits_{\footnotesize i \in 0.7<R<1} |\vec{\Pt}^i|.
\label{eq:sc10}
\end{equation}
~\\[-15pt]

The exact values of cut parameters, i.e. $\Pt^{isol}_{CUT}$,
$\epsilon^{\gamma}_{CUT}$, $\epsilon^{jet}$, $\Pt^{clust}_{CUT}$, 
$\Pt^{out}_{CUT}$, will be specified bellow since they may be
different, for instance, for various $\Pt^{\gamma}$-intervals
(more loose for higher  $\Pt^{\gamma}$).

Three criteria, 6, 7 and 9, are new and have
not been used in previous experiments. 
Their efficiency ,as well as an efficiency of other selection criteria
from the list above to reduce the background, was demonstrated in detail 
in papers \cite{BKS}, where to we refer a reader for more information.

%
\section{BACKGROUND SUPPRESSION}
%

To estimate the background for the signal events, we have done a simulation
basing on PYTHIA~5.7 (default CTEQ2L parametrization of structure functions
is used here) of
a mixture of all existing in PYTHIA QCD and SM subprocesses with large
cross sections (namely, 11--20, 28--31, 53, 68) together with our subprocesses
(2) and (3) (14 and 29 in PYTHIA). Three generations (each of $50$x$10^6$ events)
with different values of minimal $\Pt$ of hard process $\hat{p}_{\perp}^{\;min}$ 
(CKIN(3) parameter in PYTHIA) were done. 
The first one is with $\pth$ = 40 $GeV/c$, the second one is with 
$\pth$ = 100 $GeV/c$ and the third -- with $\pth$ = 200 $GeV/c$
The produced  photons were classified according to
their origin, i.e. those that are direct ones and those that result due to the 
radiation from quarks (denoted as ``$\gamma-brem$'') 
and from $\eta$-, $\omega$-, $K_S^0$-mesons decays (``$\gamma-mes$''). 
Another sort of background is formed by electrons $e^{\pm}$'s. 
\\[-22pt]
\begin{table}[h]
\begin{center}
\caption{\normalsize Number of signal and background events remained after cuts.}
\vskip-.0cm
\begin{tabular}{||c|c||c|c|c|c|c|c|c||}                  \hline \hline
\label{tab:sb1e}
$\pth$& & $\gamma$ & $\gamma$ &\multicolumn{4}{c|}{  photons from the mesons}  &
\\\cline{5-8}
\Gvc& Cuts&\hmm direct\hmm &\hmm brem\hmm & $\;\;$ $\pi^0$ $\;\;$ &$\quad$ $\eta$ $\quad$ &
$\omega$ &  $K_S^0$ &\hmm $e^{\pm}$\hmm \\\hline \hline
    &Preselected&\hmm 7795&\hmm 12951& 104919& 41845& 10984& 15058&\hmm 4204  \\\cline{2-9}
 40 &After cuts &\hmm 464&\hmm 43&     15&     0&     0&     0&\hmm   0\\\cline{2-9}
    &+ jet isol.  &\hmm 109&\hmm 7&      2&     0&     0&     0&\hmm   0\\\hline  \hline
    &Preselected&\hmm 19359  &\hmm 90022 &658981 &247644 &69210  &85568 &\hmm 47061\\\cline{2-9}
100 &After cuts&\hmm 1061 &\hmm 31 & 9 &0 &0  &0 &\hmm 3 \\\cline{2-9}   
    &+ jet isol. &\hmm 615 &\hmm14 &4 &0 &0  &0 &\hmm 2 \\\hline \hline
    &Preselected&\hmm 32629 &\hmm 207370 &780190 &288772 &82477 &98015 &\hmm 89714\\\cline{2-9}
200 &After cuts&\hmm 967 &\hmm 16& 2 &0 &0  &0 &\hmm 2\\\cline{2-9}
    &+ jet isol. &\hmm 825 &\hmm 14& 1 &0 &0  &0 &\hmm 1\\\hline \hline
\end{tabular}
\vskip0.2cm
\caption{\normalsize Efficiencies and significance values in events without jet isolation cut. }
\vskip0.1cm
\begin{tabular}{||c||c|c|c|c|c|c||}   \hline \hline
\label{tab:sb2e}
$\pth$\Gvc& $S$ & $B$ & $Eff_S(\%)$  & $Eff_B(\%)$  & $S/B$& $S/\sqrt{B}$ \\\hline \hline
40  &  464& 58 & 5.95 $\pm$ 0.28 & 0.031 $\pm$ 0.004&  8.0 & 60.9 \\\hline
100 & 1061& 43 & 5.48 $\pm$ 0.17 & 0.004 $\pm$ 0.001& 24.7 & 161.8 \\\hline  
200 & 967& 20 & 2.96 $\pm$ 0.10 & 0.002 $\pm$ 0.000& 48.4 & 216.2 
\\\hline \hline
\end{tabular}
\vskip0.2cm
\caption{\normalsize Efficiencies and significance values in events with jet isolation cut.}
\vskip0.1cm
\begin{tabular}{||c||c|c|c|c|c|c||}  \hline \hline
\label{tab:sb3e}
$\pth$\Gvc& ~~~$S$~~~ & $B$ & $Eff_S(\%)$ & $Eff_B(\%)$  & $S/B$& $S/\sqrt{B}$
 \\\hline \hline
40  & 109&  9 & 1.40 $\pm$ 0.13 & 0.005 $\pm$ 0.002& 12.1 & 36.3 \\\hline 
100 & 615& 20 & 3.18 $\pm$ 0.13 & 0.003 $\pm$ 0.000& 30.8 & 137.5 \\\hline  
200 & 825& 16 & 2.53 $\pm$ 0.09 & 0.002 $\pm$ 0.000& 51.6 & 206.3 
\\\hline \hline
\end{tabular}
\end{center}
\end{table}
\vskip-6mm
However, we also should take into account the real behavior of processes 
in the detectors. For this aim 
we have performed a detailed study (based on CMSIM GEANT simulation using 5000
generated decays of each source meson)
of difference between ECAL profiles of photon showers
from mesons and those from direct photons for $\Pt^\gamma=40\div 100~ GeV/c$ .
It has shown that the suppression factor of $\eta$-, $\omega$-,
$K_S^0$-mesons larger than 0.90 can be achieved with a selection efficiency 
of single photons
taken as 90$\%$. As for the photons from $\pi^0$ decays, the analogous 
estimations of the rejection factors
were done for the Endcap  \cite{Bar}, \cite{ECAL} and Barrel \cite{Bor}, \cite{ECAL}
 CMS ECAL regions.
They are of the order of 0.20 -- 0.70  for Barrel 
and 0.51 -- 0.75 for Endcap, depending on $\Pt^\gamma$ and a bit on 
$\eta^{\gamma}$,
for the same single photon selection efficiency 90$\%$.
Following  ~\cite{TRA}, for our estimation needs we accept the electron 
track finding efficiency to be, on the average, equal to ~$85\%$ for 
$\Pt^e\geq 40 ~GeV/c$, neglecting its $\eta$ dependence.
The number of events, selected after cuts 1 -- 9 is presented in Table 1 
(with an account of the rejection efficiencies given above) separately for signal
direct photon events and those caused by the background photons and electrons $e^{\pm}$.
Here the line ``Preselected'' corresponds to the following set of cuts:\\[-8pt]
\begin{equation}
~ \Pt^{\gamma}\geq 40 ~GeV/c, ~~~|\eta^{\gamma}|\leq 2.61,
~~ ~ \Pt^{jet}\geq 30 ~GeV/c,~~ \Pt^{hadr}\!<5 ~GeV/c,
\label{l1}
\end{equation}
according to selection rules (1), (3a).
The line ``After cuts'' contains the number of signal and background events
after selection cuts 1 -- 8 with the
values of cuts chosen as (in addition to those in point ``Preselected''):\\[-8pt]
\begin{equation}
\Pt_{CUT}^{isol}=2~GeV/c, ~~\epsilon^{\gamma}_{CUT}=5\%,
~~\dphi =15^\circ,~~\Pt^{clust}_{CUT}=10~GeV/c,~~\Pt^{out}_{CUT}=10~GeV/c.
\label{l2}
\end{equation}
~\\[-10mm]

The line ``+jet isolation'' corresponds to the complementary cut 9 of the
previous section.

The corresponding efficiencies and significance are presented in Tables 2
and 3. In these Tables
the column $S(B)$ contains the number of signal (background) events with account of 
the efficiencies described above.
$Eff_{S(B)}$ includes the values of cut efficiencies 
\footnote{taken as a ratio of the number of signal $S$
(background $B$) events, that survived cuts 1 -- 8 or 1 -- 9 from 
Section 2, to the number of the preselected events.}
and their errors.

From Table 2 it is seen that ratio $S/B$ grows  while
the \ptg~ value growing from  8.0  at $\Pt^{\gamma}\geq 40 ~GeV/c$ up to
48.4 at $\Pt^{\gamma}\geq 200 ~GeV/c$.
The jet isolation requirement (Table 3) sufficiently
improves the situation at low $\Pt$. In that case $S/B$ changes
up to 12.1 at $\Pt^{\gamma}\geq 40 ~GeV/c$ (and up to 30.8
at $\Pt^{\gamma}\geq 100 ~GeV/c$). 
As it is also seen from Tables 1 and 2  the background events admixture 
becomes nonessential for $\Pt^{\gamma} \geq 100\, GeV/c$.

The dependence of the number of events and $S/B$ ratio on two the most 
powerful cuts 
$\Pt^{out}_{CUT}$ and $\Pt^{clust}_{CUT}$ were studied in  \cite{BKS}.

In Table 4 the percentage of  
``Compton-like" process  $q~ g\rrr \gamma +q$ (as the dominant contribution
comparing with ``annihilation'' process $q~ \bar{q}\rrr \gamma + g$) events 
selected with conditions 1 -- 6 ($\Pt^{clust}_{CUT}=10~GeV/c$) is
shown for different $\Ptg$ and $\eta$ intervals: Barrel (HB) part ($|\eta|<1.4$) and
Endcap+Forward (HE+HF) part ($1.4<|\eta|<5.0$).\\[-10pt]
\setcounter{table}{4}
\begin{table}[h]
\begin{center}
\footnotesize{Table 4.}
\normalsize
\vskip.1cm
\begin{tabular}{||c||c|c|c|}                  \hline \hline
\label{tab1}
Calorimeter& \multicolumn{3}{c|}{$\Pt^{Jet}$ interval ($GeV/c$)} \\\cline{2-4}
    part   & 40--50 & 100--120 & 200--240   \\\hline \hline
HB         & 89     &  84   &  78  \\\hline
HE+HF      & 86     &  82   &  74  \\\hline 
\end{tabular}
\end{center}
\end{table}
\normalsize
~\\[-10mm]

The rates of only $q~ g\rrr \gamma +q$ events selected with conditions 1 -- 9 
 are presented for integrated luminosity
$L=100 \;pb^{-1}$ (one day of data taking at low luminosity
$L=10^{33} \,cm^{-2}s^{-1}$)
in Table 5 for different intervals of $\Pt^{\gamma}$ and
parton $x$ values. 

Cut conditions 1 -- 9, as it was shown in  \cite{BKS},
allow to select the events with a good \ptgj balance because they effectively
provide a good initial and final state radiation suppression,
i.e. suppression of the next-to-leading order diagrams.

Table 6 gives analogous values of distribution of 
the number of events in the process with a charm quark  \cite{DMS}, \cite{MD2}
$g\,c\to \gamma^{dir} \,+\,c$.
 For these tables $\Pt^{clust}_{CUT}$ was fixed to be 
$5~GeV/c$; $\Pt^{out}$ was not limited. All other cuts were put as in points
1 -- 8 of Section 2 and with cut parameter values given by \ref{l1} and \ref{l2}.
The simulation of the process $g\,b\to \gamma^{dir}\,+\,b$ $\;$ has shown
that for $b$-quark the rates are by 8 -- 10 times smaller than those for
$c$-quark.\\[-5mm]
\begin{figure}[h]
   \vskip-100mm
   \hspace{-1mm} \includegraphics[width=.52\linewidth,height=9.8cm,angle=0]{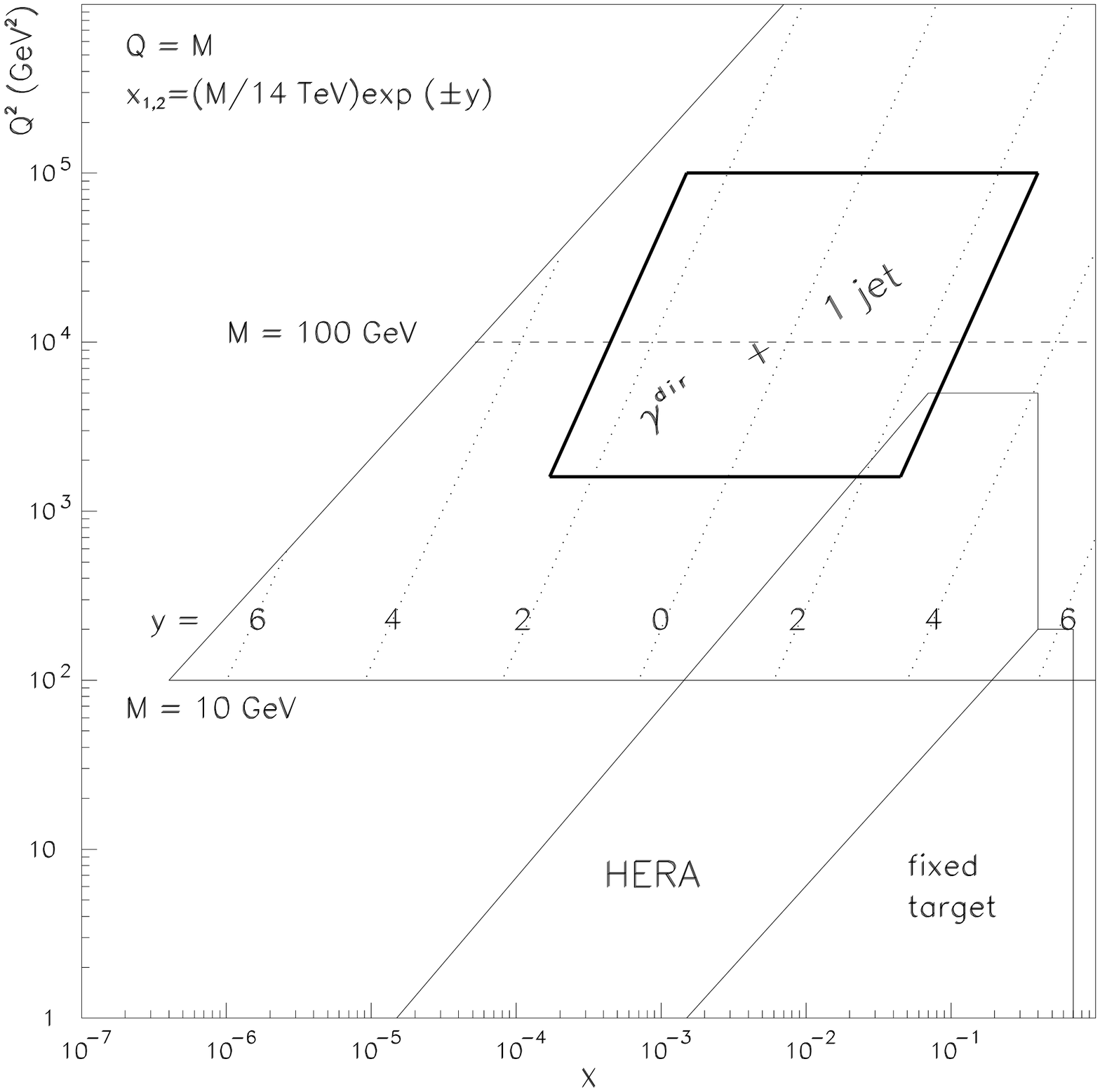}
\end{figure}
\vspace*{1.0cm}
\hspace*{8.3cm}
\parbox[r]{.45\linewidth}
{ Figure 1 shows in the widely used $(x,Q^2)$
kinematical plot (see \cite{Sti} and also in  \cite{Hu2})
what area can be covered by studying the process $q~ g\rrr \gamma +q$.
The distribution of events inside this area is given in Table 5. 
From this Figure and Table 5 it becomes clear that even at low LHC luminosity 
it would be possible to study 
the gluon distribution on a good statistics of \gpj events
in the region of small $x$ 
at values of $Q^2$ that are about 2--3 orders of magnitude higher than
those that are reached at HERA now.
It is worth emphasizing that an extension of experimentally reachable 
region at LHC to the region of lower values of $Q^2$, overlapping with the area 
covered by HERA, would be also of a big interest.}

{\vspace*{-1.3cm}
\hspace*{-.5cm} {Figure 1: \footnotesize {LHC  $(x,Q^2)$ kinematical region for
$pp\to \gamma+Jet$ process}}
}
\begin{table}[h]
\begin{center}
\caption{\normalsize Number of~ $g\,q\to \gamma^{dir} \,+\,q$~
events at different $Q^2$ and $x$ values for $L_{int}=100~pb^{-1}$}
\label{tab4}
\vskip0.1cm
\begin{tabular}{|lc|c|c|c|c|c|c|c|}                  \hline
 & $Q^2$ &\multicolumn{4}{c|}{ \hspace{-0.9cm} $x$ values of a parton} &All $x$ 
&$\Pt^{\gamma}$   \\\cline{3-7}
 & $(GeV/c)^2$ & $10^{-4}$--$10^{-3}$ & $10^{-3}$--$10^{-2}$ &$10^{-2}$--
$10^{-1}$ & $10^{-1}$--$10^{0}$ & $10^{-4}$--$10^{0}$&$(GeV/c)$     \\\hline
&\hmm\hmm 1600-2500\hmm  & 3105$\pm$104 &9715$\pm$186 &9243$\pm$183 &941$\pm$60 &23004 & 40--50\\\hline
&\hmm\hmm 2500-5000\hmm  & 1217$\pm$52 & 5539$\pm$100 & 5794$\pm$102 &930$\pm$36 &13481 & 50--71\\\hline
&\hmm\hmm 5000-10000\hmm & 144$\pm$9 & 1502$\pm$29 & 1671$\pm$30 & 407$\pm$14 & 3724   & 71--100\\\hline
&\hmm\hmm 10000-20000\hmm& 6$\pm$1 & 328$\pm$8 &  422$\pm$9 & 161$\pm$5 & 916  &  100--141\\\hline
&\hmm\hmm 20000-40000\hmm& 0 & 65$\pm$2 &102$\pm$2& 52$\pm$2 & 219 & 141--200              \\\hline
&\hmm\hmm 40000-80000\hmm& 0 & 9$\pm$1& 18$\pm$1& 11$\pm$1 &  37 & 200--283       \\\hline
\end{tabular}
\vskip0.3cm
\caption{\normalsize Number of~ $g\,c\to \gamma^{dir} \,+\,c$~ events at different $Q^2$ and $x$ values for $L_{int}=100~pb^{-1}$}
\label{tab5}
\vskip0.1cm
\begin{tabular}{|lc|c|c|c|c|c|c|c|}                  \hline
& $Q^2$ &\multicolumn{4}{c|}{ \hspace{-1.2cm} $x$ values for $c$-quark} & All $x$ 
& $\Pt^{\gamma}$ \\\cline{3-7}
& $(GeV/c)^2$ & $10^{-4}$--$10^{-3}$ & $10^{-3}$--$10^{-2}$ &$10^{-2}$--
$10^{-1}$ & $10^{-1}$--$10^{0}$ & $10^{-4}$--$10^{0}$&$(GeV/c)$  \\\hline
&\hmm\hmm 1600-2500\hmm &  426$\pm$39 & 1395$\pm$71 & 1495$\pm$73 & 161$\pm$24 & 3477 & 40--50   \\\hline
&\hmm\hmm 2500-5000\hmm &  155$\pm$18 & 806$\pm$39 &841$\pm$39 &86$\pm$11 & 1888 & 50--71  \\\hline
&\hmm\hmm 5000-10000\hmm &  18$\pm$3 & 214$\pm$11 & 244$\pm$12 & 50$\pm$5 & 526 & 71--100\\\hline
&\hmm\hmm 10000-20000\hmm &  1$\pm$1 & 37$\pm$3 & 51$\pm$3 & 17$\pm$2  & 106 & 100--141      \\\hline
&\hmm\hmm 20000-40000\hmm &  0 & 6$\pm$1 & 14$\pm$1 & 4$\pm$0.3  & 24& 141--200    \\\hline
&\hmm\hmm 40000-80000\hmm &  0 & 1$\pm$0.2 & 2$\pm$0.2& 1$\pm$0.2 & 4& 200--283   \\\hline
\end{tabular}
\end{center}
\end{table}

\section{SUMMARY}
It is shown that the sample of \gpj events with a clean topology, 
which is most suitable for jet energy absolute scale setting at LHC
energy (selected with the cut conditions of \cite{BKS}
that powerfully suppress initial and final state radiation,
i.e. next-to-leading order diagrams contribution), covers
the kinematical region of $x$ values as small as accessible
at HERA  \cite{H1}, \cite{ZEUS}, but
at much higher $Q^2$ values (2--3 orders of magnitude): 
$10^{-4}\leq x \leq 10^{-1}$ with $1.6\cdot10^3\leq Q^2\leq10^5 ~(GeV/c)^2$.
It is shown that percentage of gluon dominated subprocess 
$q g\rrr \gamma +q$ events is about $75-90\%$ among
\gpj events 
what would allow, in principle, a good extraction of gluon
distribution function from future LHC \gpj data.

~\\[3mm]
\noindent
{\bf Acknowledgements}
\newline
We thank P.~Aurenche, D.~Denegri, M.Dittmar, M.~Fontannaz, J.Ph.~Guillet, M.L.~Mangano, 
E.~Pilon and S.~Tapprogge for helpful discussions.

\end{document}